\documentclass[12pt,draft]{article}

\usepackage{amssymb}
\usepackage{latexsym}

\newtheorem{theorem}{Theorem}[section]
\newtheorem{definition}{Definition}[section]
\newtheorem{lemma}{Lemma}[section]
\newtheorem{proposition}[theorem]{Proposition}

\newcommand{\nc}{\newcommand}
\nc{\stack}[2]{{\begin{array}{c}
\scriptstyle #1 \\ \scriptstyle #2 \end{array}} }
\nc{\C}{{\mathbb C}}
\nc{\R}{{\mathbb R}}
\nc{\HH}{{\mathbb H}}
\nc{\Z}{{\mathbb Z}}
\nc{\N}{{\mathbb N}}
\nc{\dd}{{\rm d}}
\nc{\DD}{{\bf d}}
\nc{\ii}{{\bf i}}
\nc{\jj}{{\bf j}}
\nc{\kk}{{\bf k}}
\nc{\co}{{\cal O}}
\nc{\tr}{\mathop{\rm tr}\nolimits}
\nc{\su}{{\mathfrak s}{\mathfrak u}(2)}
\nc{\so}{{\mathfrak s}{\mathfrak o} (4)}

\begin{document} 

\title{Classification of 't Hooft instantons over\\
multi-centered gravitational instantons} 

\author{
G\'abor Etesi
\\ {\it R\'enyi Institute of Mathematics,}
\\ {\it  Hungarian Academy of Sciences,}
\\{\it Re\'altanoda u. 13-15,} 
\\{\it H-1053 Budapest, Hungary} 
\\ {\tt etesi@math-inst.hu}}

\maketitle

\pagestyle{myheadings}
\markright{G. Etesi: Classification of instantons over Gibbons--Hawking
spaces}  

\thispagestyle{empty}

\begin{abstract}
This work presents a classification of all
smooth 't Hooft--Jackiw--Nohl--Rebbi instantons over Gibbons--Hawking
spaces. That is, we find all smooth $SU(2)$ Yang--Mills instantons over
these spaces which arise by conformal rescalings of the metric with
suitable functions. 

Since the Gibbons--Hawking spaces are hyper-K\"ahler gravitational
instantons, the rescaling functions must be positive harmonic. 
By using twistor methods we present integral formulae for the kernel of
the Laplacian associated to these spaces. These integrals are
generalizations of the classical Whittaker--Watson formula. By the aid 
of these we prove that all 't Hooft instantons have 
already been found in a recent paper \cite{ete-hau3}. 

This result also shows that actually all such smooth 't
Hooft--Jackiw--Nohl--Rebbi instantons describe singular magnetic
monopoles on the flat three-space with zero magnetic charge moreover the
reducible ones generate the full $L^2$ cohomolgy of the Gibbons--Hawking
spaces. 
\end{abstract}
\vspace{0.5cm}
\centerline{PACS numbers: 02.40.Ma, 02.30.Jr, 11.15.Tk, 04.20.Jb}
\centerline{Keywords: {\it Gravitational and Yang--Mills
instantons, harmonic functions, twistors}}

\section{Introduction}
The simplest non-flat geometries in four dimensions are
provided by non-compact hyper-K\"ahler spaces. These spaces also appear
naturally in recent investigations in high energy physics. For
example, physicists call these spaces as ``gravitational instantons''
because hyper-K\"ahler metrics are Ricci flat hence solve Einstein's
equatios and have self-dual curvature tensor \cite{gib-haw}. 

Moreover, on the one hand, motivated by Sen's S-duality conjecture
\cite{sen} from 1994, recently there have been some interest in 
understanding the $L^2$ cohomology of certain moduli spaces of magnetic
monopoles carrying natural hyper-K\"ahler metrics. Probably the 
strongest evidence in Sen's paper for his conjecture was
an explicit construction of an $L^2$ harmonic $2$-form on the universal 
double cover of the Atiyah--Hitchin manifold. Sen's conjecture also
predicted an $L^2$ harmonic $2$-form on the Euclidean Taub--NUT space. 
This was found later by Gibbons in 1996 \cite{gib}. He constructed
it as the exterior derivative of the dual of the Killing field of a
canonical $U(1)$ action. In a joint paper with T. Hausel we imitated
Gibbons' construction yielding a classification of $L^2$ harmonic forms on
the Euclidean Schwarzschild manifold \cite{ete-hau1}.
Recently Hausel, Hunsicker and Mazzeo classified the $L^2$ harmonic forms
over various spaces with prescribed infinity, including gravitational
instantons \cite{hau-hun-maz}. 

On the other hand, it is natural to ask wether or not the ADHM
construction, originally designed for finding
Yang--Mills instantons over flat $\R^4$, is extendible for hyper-K\"ahler
spaces. Kronheimer provided an $A$-$D$-$E$-type classification of the
so-called ALE (Asymptotically Locally Euclidean) hyper-K\"ahler spaces in
1989 \cite{kro2}. Later Cherkis and Kapustin went on and constructed
$A$-$D$ sequences of so-called ALF (Asymptotically Locally Flat) and five
$D_4$ type ALG (this shortening is by induction) gravitational instantons
\cite{che-kap1}\cite{che-kap2}. These are also non-compact hyper-K\"ahler
spaces with different asymptotical behaviour. The question naturally
arises what about the ADHM construction over these new gravitational
instantons.

Kronheimer and Nakajima succeeded to extend the ADHM construction
to ALE gravitational instantons yielding a classification of Yang--Mills
instantons over these spaces in 1990 \cite{kro-nak}. In the above scheme
the Gibbons--Hawking gravitational instantons appear as the $A_k$ ALE and
$A_k$ ALF spaces and are distinguished because  the metrics are
known explicitly on them. In our papers \cite{ete-hau2}\cite{ete-hau3} we
could construct $SU(2)$ Yang--Mills instantons over these spaces via the
conformal rescaling method of Jackiw--Nohl--Rebbi \cite{jac-noh-reb}.
Moreover it turned out that finding the reducible instantons among our
solutions we could describe all the $L^2$ harmonic forms on them in a
natural geometric way yielding a kind of unification of the two above,
apparently independent problems.

In this paper, as a completion of these investigations we prove that we
have already found all instantons which can be constructed by the
conformal rescaling method \cite{ete-hau3}. The proof is based on a
standard twistor theoretic construction of the kernel of the Laplacian. 

The paper is organized as follows. In Sec. 2 we recall the
't Hooft--Jackiw--Nohl--Rebbi construction in the form of the
Atiyah--Hitchin--Singer theorem \cite{ati-hit-sin}. This enables us to
conclude that the scaling functions we are seeking must be harmonic
with respect to the Gibbons--Hawking geometries. 

In Sec. 3 we construct all solutions of the Laplace equation over
$\C^2\setminus\{ 0\}/\Z_N$ via a classical harmonic
expansion. This is possible because the metric collapsed to this
space possesses a large $U(2)$ isometry. From here we can see that only
those harmonic functions give rise to everywhere non-singular instantons
which are positive everywhere with at most pointlike singularities.

In Sec. 4 we perturb the above singular solutions into regular
ones by constructing integral formulae for harmonic functions
in the non-singular case based on twistor theory. These
integral formulae are generalizations of the classical
Whittaker--Watson formula over flat $\R^3$ \cite{whi-wat}. 

In light of these investigations we can conclude that all 't Hooft
instantons have in fact been found in our earlier paper \cite{ete-hau3}.
Since all these functions are invariant under the natural $U(1)$ action on
the Gibbons--Hawking spaces the resulting smooth Yang--Mills instantons
describe singular magnetic monopoles on the flat $\R^3$, as 
it was pointed out by Kronheimer in his diploma thesis \cite{kro1}. We
also can see that the reducible solutions generate the full $L^2$
cohomology of the spaces in question. All these theorems are collected in
Sec. 5. 

Finally, we take conclusion and outlook in Sec. 6 and argue that our
methods work for the above mentioned "exotic" gravitational instantons,
too.

\section{The Atiyah--Hitchin--Singer theorem}

First we recall the general theory from \cite{ati-hit-sin}.
Let $(M,g)$ be a four-dimensional Riemannian spin-manifold. Remember that
via Spin$(4)\cong SU(2)\times SU(2)$ we have a Lie algebra isomorphism
$\so\cong\su^+\oplus\su^-$. Consider the Levi--Civit\'a connection
$\nabla$ which is locally represented in a fixed gauge by an
$\so$-valued 1-form $\omega$ on $TU$ for $U\subset M$. Because $M$ is spin
and four-dimensional, we can consistently lift
this connection to the spin connection $\nabla_S$, locally given by
$\omega_S$, on the spin bundle $SM$ (which is a complex bundle of rank
four) and can project it to the $\su^\pm$ components denoted by
$\nabla^\pm$. The projected connections loacally are given by
$\su^\pm$-valued 1-forms $A^\pm$ and live on the chiral spinor bundles
$S^\pm M$ where the decomposition $SM=S^+M\oplus S^-M$ corresponds to the
above splitting of Spin$(4)$. One can raise the question what are the
conditions on the metric $g$ for either $\nabla^+$ or $\nabla^-$ to be
self-dual (seeking antiself-dual solutions is only a matter of
reversing the orientation of $M$).

Consider the curvature 2-form $R\in C^\infty (\Lambda^2M\otimes\so )$ of
the metric. There is a standard linear isomorphism
$\so\cong\Lambda^2\R^4$ given by $A\mapsto\alpha$ with $xAy=\alpha
(x,y)$ for all $x,y\in\R^4$. Therefore we may regard $R$ as a {\it
2-form-valued 2-form}
in $C^\infty (\Lambda^2M\otimes \Lambda^2M)$ i.e. for vector fields $X,Y$
over $M$ we have $R(X, Y)\in C^\infty (\Lambda^2M)$. Since the space
of four dimensional curvature tensors, acted on by $SO(4)$, is 20
dimensional, one gets a 20 dimensional
reducible representation of $SO(4)$ (and of Spin$(4)$, being $M$
spin). The decomposition into irreducible components is (see \cite{bes},
pp. 45-52)
\begin{equation}
R={1\over 12}\pmatrix{s & 0\cr 0 & s\cr}+\pmatrix{0 &
B\cr B^T & 0\cr}+
\pmatrix{W^+ & 0\cr 0 & 0\cr}+\pmatrix{0 & 0\cr 0 & W^-\cr},
\label{gorbuletszetszedes}
\end{equation}
where $s$ is the scalar curvature, $B$ is the traceless Ricci tensor,
$W^\pm$ are the Weyl tensors. The splitting of the Weyl tensor is a
special four-dimensional phenomenon and is related with the above 
splitting of the Lie algebra $\so$. There are two Hodge operations which
can operate on $R$. One (denoted by $*$) acts on the {\it 2-form part
of $R$} while the other one (denoted by $\star$) acts on the {\it values}
of $R$ (which are also 2-forms). In a local coordinate system, these
actions are given by
\[(\star R)_{ijkl}={1\over
2}\sqrt{\det g}\:\varepsilon_{ijmn}R^{mn}_{\:\:\:\:\:\:kl},\]
\[(*R)_{ijkl}={1\over
2}\sqrt{\det g}\:R_{ij}^{\:\:\:\:mn}\varepsilon_{mnkl}.\]
It is not difficult to see that the projections $p^\pm
:\so\rightarrow\su^\pm$ are given by $R\mapsto F^\pm:={1\over 2}(1\pm\star
R)$, and $F^\pm$ are self-dual with respect to $g$ if
and only if $*(1\pm\star R)=(1\pm\star R)$. Using the
previous representation for the decomposition of $R$ suppose $\star$ acts
on the left while $*$ on the right, both of them via
  \[\pmatrix{{\rm id} & 0\cr 0 & {\rm -id}\cr}.\]
In this case the previous self-duality condition looks like
($\overline{W}^\pm :=W^\pm+{1\over 12}s$)
\[\pmatrix{\overline{W}^+\pm\overline{W}^+ & -(B\pm B)\cr B^T\mp B^T&
-\left(\overline{W}^-\mp\overline{W}^-\right)\cr}=   
\pmatrix{\overline{W}^+\pm\overline{W}^+
&B\pm B\cr B^T\mp B^T & \overline{W}^-\mp\overline{W}^-\cr}.\]
From here we can immediately conclude that $F^+$ is self-dual if and
only if $B=0$ i.e. $g$ is {\it Einstein} while $F^-$ is self-dual if
and only if $\overline{W}^-=0$ i.e. $g$ is {\it
half-conformally flat (i.e. self-dual) with vanishing scalar
curvature}. Hence we have proved \cite{ati-hit-sin}:
\begin{theorem} [Atiyah--Hitchin--Singer] Let
$(M,g)$ be a four-dimensional Riemannian spin manifold. Then 

\noindent {\rm (i)} $F^+$ is the curvature of
an self-dual $SU(2)$-connection on $S^+M$ if and only if $g$ is Einstein,
or

\noindent {\rm (ii)} $F^-$ is the curvature of a self-dual
$SU(2)$-connection on $S^-M$ if and only if $g$ is half
conformally flat (i.e. self-dual) with vanishing scalar
curvature. $\Diamond$
\end{theorem}

Remember that both the (anti)self-duality equations
\[*F=\pm F\]
and the action
\[\Vert F\Vert^2={1\over 8\pi^2}\int\limits_M\vert
F\vert^2_g=-{1\over 8\pi^2}\int\limits_M\tr (F\wedge *F)\]
are conformally invariant in four dimensions; consequently if we can
rescale $g$ with a suitable  function $f$ producing a metric
$\tilde g$ which satisfies one of the properties of the previous theorem
then we can construct instantons over the original manifold $(M,g)$. This
idea was used by Jackiw, Nohl and Rebbi to construct instantons over the
flat $\R^4$ \cite{jac-noh-reb}. Consequently we find it convenient to
make the following
\begin{definition}
Let $(M,g)$ be a four dimensional Riemannian spin manifold and
$\nabla^\pm$ be a smooth self-dual $SU(2)$ connection of finite action on
the chiral spinor bundle $S^\pm M$. If there is a smooth function
$f: M\rightarrow\R$ such that the projected Levi--Civit\'a
connection of $f^2g$ is gauge equivalent to $\nabla^\pm$ then this
connection is called an {\em 't Hooft instanton} over $(M,g)$.
\end{definition}

First consider the case of $F^+$, i.e. part (i) of the above theorem
\cite{ete-hau2}. Let 
$(M,g)$ be a Riemannian manifold of dimension $n>2$. Remember that $\psi
:M\rightarrow M$ is a {\it conformal isometry} of
$(M,g)$ if there is a function $f: M\rightarrow\R$ such that
$\psi^*g=f^2g$. Notice that being $\psi$ a diffeomorphism, $f$ cannot be
zero anywhere i.e. we may assume that it is positive, $f>0$. Ordinary
isometries are the special cases with $f=1$. The
vector field $X$ on $M$, induced by the conformal isometry, is
called a {\it conformal Killing field}. It satisfies the {\it conformal
Killing equation} (\cite{wal}, pp. 443-444) 
\[L_Xg-{2{\rm div}X\over n}g=0\]
where $L$ is the Lie derivative while div is the divergence of a
vector field. If $\xi =\langle X,\:\cdot\:\rangle$ denotes the dual 1-form
to $X$ with respect to the metric, then consider the following {\it
conformal Killing data}:
\begin{equation}
(\xi ,\:\:\:\dd\xi ,\:\:\:{\rm div}X,\:\:\: \dd {\rm div}X).
\label{adatok}
\end{equation}
These satisfy the following equations (see \cite{gar}):
\begin{equation}
\begin{array}{ll}
\nabla\xi =(1/2)\dd\xi +(1/n)({\rm div}X)g,\\
\\
\nabla (\dd\xi )=(1/n)(g\otimes\dd {\rm div}X -(\dd{\rm 
div}X)\otimes g)+2R(\:\cdot\:,\:\cdot\:,\:\cdot\:,\xi ),\\
\\
\nabla ({\rm div}X)=\dd{\rm div}X,\\
\\
\nabla (\dd {\rm div}X)=-(n/2)\nabla_XP-({\rm
div}X)P-(n/2)Q.\\
\end{array}
\label{egyenletek}
\end{equation}
Here $R$ is understood as the $(3,1)$-curvature tensor while
\[P=r-{s\over (n-1)(n-2)}g,\:\:\:\:\:Q={\rm tr}(P\otimes\dd\xi
+\dd\xi\otimes P)\] 
with $r$ being the
Ricci-tensor. For clarity we
remark that $Q_{ij}=(P^k_i\dd\xi_{jk}+P^k_j\dd\xi_{ik})
$ i.e., tr is the only non-trivial contraction.
If $\gamma$ is a smooth curve in $M$ then fixing conformal Killing data in
a point $p=\gamma (t)$ we can integrate (\ref{egyenletek}) to get all the
values of $X$ along $\gamma$. Actually if $X$ is a conformal Killing field
then by fixing the above data in one point $p\in M$ we can determine the
values of $X$ over the {\it whole} $M$, provided that $M$ is connected. 
Consequently, if these data vanish in one point, then $X$ vanishes over
all the $M$. 

Furthermore a Riemannian manifold $(M,g)$ is called {\it irreducible} if 
the holonomy group, induced by the metric, acts irreducibly on each
tangent space of $M$. 

Now we can state \cite{ete-hau2}:

\begin{proposition} 
Let $(M,g)$ be a connected, irreducible, Ricci-flat Riemannian
manifold of dimension $n>2$. Then $(M,\tilde{g})$ with
$\tilde{g}=\varphi^{-2}g$ is Einstein if and only if $\varphi$ is a
non-zero constant function on $M$. 
\end{proposition}

\noindent {\it Remark.} Notice that the above proposition is not true
for reducible manifolds: the already mentioned Jackiw--Nohl--Rebbi
construction \cite{jac-noh-reb} provides us with non-trivial Einstein metrics,
conformally equivalent to the flat $\R^4$. 
\vspace{0.1in}

\noindent {\it Proof}. If $\eta$ is a 1-form on $(M,g)$ with
dual vector $Y=\langle\eta ,\:\cdot\:\rangle$, then the $(0,2)$-tensor
$\nabla\eta$ can be decomposed into antisymmetric, trace and traceless
symmetric parts respectively as follows (e.g. \cite{pet}, p. 200,
Ex. 5.):
\begin{equation}
\nabla\eta={1\over 2}\dd\eta -{\delta\eta\over n}g+{1\over 2}\left(
L_Yg+{2\delta\eta\over n}g\right)
\label{dekompozicio}
\end{equation}
where $\delta$ is the exterior codifferentiation on $(M,g)$
satisfying $\delta\eta =-{\rm div}Y$. Being $g$ an Einstein metric, it
has identically zero traceless Ricci tensor i.e. $B=0$ from 
decomposition (\ref{gorbuletszetszedes}). We rescale
$g$ with the function $\varphi :M\rightarrow\R^+$ as
$\tilde{g}:=\varphi^{-2}g$. Then the traceless Ricci part of the new
curvature is (see \cite{bes}, p. 59)
\[\widetilde{B}= {n-2\over\varphi}\left(\nabla^2\varphi
+{\triangle\varphi\over n}g\right) .\]  
Here $\triangle$ denotes the Laplacian with respect to $g$. From here we
can see that if $n>2$, the condition for $\tilde{g}$ to be again Einstein
is
\[\nabla^2\varphi +{\triangle\varphi\over n}g =0.\]
However, if $X:=\langle\dd\varphi ,\:\cdot\:\rangle$ is the dual vector
field then we can write by (\ref{dekompozicio}) that
\[\nabla^2\varphi ={1\over 2}\dd^2\varphi -{\delta (\dd\varphi
)\over n}g+{1\over 2}\left( L_Xg+{2\delta (\dd\varphi )\over n}g\right) =
-{\triangle\varphi\over n}g+{1\over 2}\left( L_Xg+{2\triangle\varphi\over 
n}g\right).\]
We have used $\dd^2=0$ and $\delta\dd =\triangle$ for functions. Therefore
we can conclude that $\varphi^{-2}g$ is Einstein if and only if
\[L_Xg+{2\triangle\varphi\over n}g=L_Xg-{2{\rm div}X\over n}g=0\]
i.e. $X$ is a conformal Killing field on $(M,g)$ obeying 
$X=\langle\dd\varphi ,\:\cdot\:\rangle$. The conformal Killing data
(\ref{adatok}) for this $X$ are the following:
\begin{equation}
(\dd\varphi ,\:\:\:\dd^2\varphi =0,\:\:\: -\triangle\varphi ,\:\:\:-\dd
(\triangle\varphi )).
\label{konkretadatok}
\end{equation}
Now we may argue as follows: the last equation of
(\ref{egyenletek}) implies that 
\[\nabla ( \dd (\triangle\varphi ))=0\]
over the Ricci-flat $(M, g)$. By virtue of the irreducibility of
$(M,g)$ this means that actually $\dd (\triangle\varphi )=0$ (cf.
e.g. \cite{bes}, p. 282, Th. 10.19) and hence $\triangle\varphi
=$const. over the whole $(M,g)$. Consequently, the second equation of
(\ref{egyenletek}) shows that for all $Y,Z,V$ we have
\[R(Y,Z,V, \dd\varphi )=0.\]
Taking into account again that $(M,g)$ is irreducible, there is a point
where $R_p$ is non-zero. Assume that the previous equality holds for all
$Y_p,Z_p,V_p$ but $\dd\varphi_p\not=0$. This is possible only if a
subspace, spanned by $\dd\varphi_p$ in $T^*_pM$, is invariant under the
action of the holonomy group. But this contradicts the irreducibility
assumption. Consequently $\dd\varphi_p=0$. Finally, the first equation of
(\ref{egyenletek}) yields that $\triangle\varphi (p)=0$ 
i.e. $\triangle\varphi =0$. Therefore we can conclude that in that
point all the conformal data (\ref{konkretadatok}) vanish implying $X=0$.
In other words $\varphi$ is a non-zero constant. $\Diamond$
\vspace{0.1in}

\noindent In light of this proposition, general Ricci-flat 
manifolds cannot be rescaled into Einstein manifolds in a non-trivial
way. Notice that the Gibbons--Hawking spaces (see below) are
irreducible Ricci-flat manifolds. If this was not the
case, then, taking into account their simply connectedness and geodesic
completeness, they would split into a Riemannian product 
$(M_1\times M_2, g_1\times g_2)$ by virtue of the de Rham theorem
\cite{deR}. But it is easily checked that this is not the case. We just
remark that the same is true for the Euclidean Schwarzschild manifold. 

Consequently constructing instantons in this way is not very
productive. 

\section{Construction of singular Laplace operators}
Therefore we turn our attention to the condition on the $F^-$ part of
the metric curvature in the special case of the Gibbons--Hawking spaces.

First we give a brief description of the Gibbons--Hawking spaces denoted
by $M_V$. These spaces can be understood topologically as follows
(cf., e.g. \cite{hit1}). Take an $N\in\N^+$ and
consider the cyclic group $\Z_N\subset SU(2)$ acting on $\C^2$
induced by the $SU(2)$ action. The resulting quotient $\C^2/\Z_N$ is
singular at the origin and looks like $\R\times (S^3/\Z_N)$ at infinity.
If $(v,w)\in\C^2$ then the monomials $v^N, w^N, vw$ are invariant under
$\Z_N$. If we denote them by $x,y,z$ then they satisfy an algebraic
equation 
\[xy-z^N=0.\]
This gives an isomorphism between $\C^2/\Z_N$ and the complex surface
$xy-z^N=0$ in $\C^3$. We can look at deformations of this singularity.
As it is well known, this surface is singular because the monomial $z^N$
contains multiple roots i.e., its discriminant is zero. This can be
removed by adding lower order terms in $z$ such that the resulting
polynomial in $z$ is of nonzero discriminant (``blowing up''):
\begin{equation}
xy-(z^N+a_1z^{N-1}+\dots +a_N)=0.
\label{ghter}
\end{equation}
These complex surfaces still have the required topology at infinity but
are nonsingular. We shall take $M_V$ to be the underlying
four-dimensional differentiable manifold of such a surface. 

One also can construct $M_V$ intuitively.
Start with $\R^3\times S^1$ acted on by $U(1)$ along the circles.
Consider $N$ distinct points $p_1,\dots,p_N$ along the $z$ axis of $\R^3$
for example and shrink the $S^1$ circles over them. The resulting space is
$M_V=\C^2\setminus\{ 0\}/\Z_N\cup E$ with $E$ the exceptional divisor
consisting of $N-1$ copies of $\C P^1$'s, attached to each
other according to the $A_{N-1}$ Dynkin diagram. From here we can also see
that there is a circle action  on $M_V$ with $N$ fixed points
$p_1,\dots,p_N\in M_V$, called NUTs. 
The quotient is $\R^3$ and we denote the images of the fixed points also
by $p_1,\dots ,p_N\in \R^3$. Then $U_V:=M_V\setminus\{ p_1,\dots,p_N\}$ is
fibered over $\R^3\setminus\{ p_1,\dots,p_N\}$ with $S^1$ fibers. The
degree of this circle bundle around each point $p_i$ is one. 

The metric $g_V$ on $U_V$ looks like (cf. e.g. \cite{hit1} or p. 363
of \cite{egu-gil-han})
\begin{equation}
\dd s^2=V(\dd x^2+\dd y^2+\dd z^2)+{1\over V}(\dd\tau +\alpha )^2,
\label{metrika}
\end{equation}
where $\tau\in (0,8\pi m]$ parametrizes the circles and
$x=(x,y,z)\in\R^3$; the smooth function $V: \R^3\setminus\{
p_1,\dots,p_N\}\rightarrow\R^+$ and the
1-form $\alpha\in C^\infty (\Lambda^1(\R^3\setminus\{ p_1,\dots,p_N\}))$
are defined up to gauge transformations as follows:
\[ V(x ,\tau )=V(x )=c+\sum\limits_{s=1}^N{2m\over\vert x-p_s\vert}
,\:\:\:\:\:*_3\dd\alpha =\dd V.\]
Here $c$ is a parameter with values 0 or 1 and $*_3$ refers to the
Hodge-operation with respect to the flat metric on $\R^3$. We can see
that the metric is independent of $\tau$ hence we have a Killing field on
$(U_V,g_V|_{U_V})$. This Killing field provides the above mentioned
$U(1)$-action. Furthermore it is possible to show that, despite the
apparent singularities in the NUTs, all things extend analytically over
the whole $M_V$ providing a real analytic space $(M_V, g_V)$. 

Notice that for a fixed manifold $M_V$ we have {\it two}
different metrics on it corresponding to the two possible values of $c$.
Take $c=0$ then for $N=1$ we just find the flat metric on $\R^4$
while $N=2$ gives rise to the Eguchi--Hanson space on $T^*\C P^1$. If
$c=1$ and $N=1$ we recover the Taub--NUT metric on $\R^4$. All these
spaces posses an $U(2)$ isometry. For higher $N$'s the resulting spaces
are the multi-Eguchi--Hanson and multi-Taub--NUT spaces respectively.
These metrics admit only an $U(1)$ isometry coming from the circle action
mentioned above. 

These two infinite sequences of metrics are hyper-K\"ahler
spaces i.e. for their curvature $s=0$ and $W^-=0$
holds (using another terminology they are half conformally flat with
vanishing scalar curvature). Consequently part (ii) of the
Atiyah--Hitchin--Singer theorem can be applied for them. Our aim is to
find metrics $\tilde{g}$ (as much as possible), conformally equivalent to
a fixed Gibbons--Hawking metric $g_V$, such that $\tilde{g}$'s are
self-dual and have vanishing scalar curvature: in this case the metric
instantons in $\tilde{g}$'s provide self-dual connections on them, as we
have seen. Taking into account that the $(3,1)$-Weyl tensor $W$ is
invariant under conformal rescalings i.e. $\widetilde{W}=W$, the condition
$\widetilde{W}^-+\tilde{s}/12 =0$ for the
$\tilde{g}$'s settles down for having vanishing scalar curvature
$\tilde{s}=0$. Consider the rescaling $g\mapsto\tilde{g}:=f^2g$
where $f: M_V\rightarrow\R$ is a function. In this case
the scalar curvature transforms as $s\mapsto\tilde{s}$ where $\tilde{s}$
satisfies (see \cite{bes}, pp. 58-59):
\[f^3\tilde{s}=6\triangle f +fs.\]
Taking into account that the Gibbons--Hawking spaces are Ricci-flat, our
condition for the scaling function amounts to the simple condition
\begin{equation}
\triangle f =0
\label{laplace}
\end{equation}
i.e. it must be a harmonic function (with respect to the Gibbons--Hawking
geometry). In other words we have to analyse the kernel of the Laplacian
for our aim.

To analyse harmonic functions in these geometries, first we construct
Laplacians on the collapsed spaces $\C^2\setminus\{ 0\}/\Z_N$ because
these operators and their associated harmonic functions can be constructed
explicitly.

Assume therefore that all the NUTs are pulled together in
(\ref{ghter}) i.e. we return to the singular quotient $\C^2/\Z_N$. If we
remove the singular origin the resulting space is topologically a lense
space: $\C^2\setminus\{ 0\}/\Z_N\cong \R^+\times (S^3/\Z_N)$. We
shall denote this space by $M_{V_0}$. 
Regarding it as a plane bundle over $S^2$ and denoting by $(\Theta ,\phi)$
the spherical coordinates while by $(r,\tau )$ polar coordinates on the
fibres we simply get
\[V_0(r)={cr+2mN\over r},\:\:\:\:\:\alpha_0 =2m\cos\Theta\dd (N\phi )\]
under this blowing down process. Consequently the Gibbons--Hawking metric
(\ref{metrika}) reduces to a metric $g_0$ of the local form
\[\dd s^2_0={cr+2mN\over r}(\dd
r^2+r^2(\dd\Theta^2+\sin^2\Theta\dd\phi^2))+{r\over
cr+2mN}(2mN)^2(\dd\tau +\cos\Theta\dd\phi )^2\]
on $M_{V_0}$ where 
\[0<r<\infty ,\:\:\:\:\:0\leq\tau
<{4\pi\over N} ,\:\:\:\:\:0\leq\phi <2\pi
,\:\:\:\:\:0\leq \Theta <\pi .\]
We remark that for $N=1$ this metric extends as the flat metric for
$c=0$ and as the Taub-NUT metric for $c=1$ over $\R^4$. 
One easily calculates $\det g_0=4m^2N^2(cr+2mN)^2r^2\sin^2\Theta$ 
and by using the local expression
\[\triangle =\sum\limits_{i,j}{1\over\sqrt{\det g}}{\partial\over\partial
x^i}\left(\sqrt{\det g}\:g^{ij}{\partial\over\partial x^j}\right) ,\]
where $g^{ij}$ are the components of the inverse matrix, the singular
Laplacian looks like
\[\triangle_0={(cr+2mN)^2\sin^2\Theta +4m^2N^2\cos^2\Theta\over 4m^2N^2
(cr^2+2mNr)\sin^2\Theta}{\partial^2\over\partial\tau^2}+{r\over
cr+2mN}{\partial^2\over\partial r^2}+{2\over cr+2mN}{\partial\over\partial
r}\]
\[+{1\over 
cr^2+2mNr}\left({\partial^2\over\partial\Theta^2}+\cot\Theta{\partial\over
\partial\Theta}\right)+{1\over
(cr^2+2mNr)\sin^2\Theta}\left({\partial^2\over\partial\phi^2}
-2\cos\Theta{\partial^2\over\partial\tau\partial\phi}\right) .\]
This apparently new Laplacian can be rewritten in old terms as follows.
The lense space with its standard metric induced from the round $S^3$
has an associated Laplacian
\[\triangle_{S^3/\Z_N} ={1\over
\sin^2\Theta}\left({\partial^2\over\partial\tau^2}-2\cos\Theta{\partial^2
\over\partial\tau\partial\varphi}+{\partial^2\over\partial\varphi^2}\right)
+{\partial^2\over\partial\Theta^2}+\cot\Theta{\partial\over\partial\Theta}\]
in Euler coordinates. By the aid of this we obtain 
\[\triangle_0={r\over cr+2mN}{\partial^2\over\partial
r^2}+{2\over
cr+2mN}{\partial\over\partial r}+{1\over cr^2 +2mNr}\triangle_{S^3/\Z_N}+
{(cr+2mN)^2-4m^2N^2\over
4m^2N^2(cr^2+2mNr)}{\partial^2\over\partial\tau^2}.\]

Now we wish to solve (\ref{laplace}) in its
full generality in this case. To this end we introduce an orthonormal
system of bounded smooth functions on $S^3/\Z_N$ by the (three
dimensional) spherical harmonics $Y^{kl}_j$ with $j=0,1,\dots$ and
$k,l=0,\dots ,j$ (for a fixed $j$, there are $(j+1)^2$ independent
spherical harmonics on $S^3/\Z_N$). In Euler coordinates these take the
shape
\[Y^{kl}_j(\tau, \phi ,\Theta )={1\over
4\pi}\sqrt{{(2j+1)(j-|k-l|)!\over (j+|k-l|)!}}\:{\rm e}^{\ii
kN\tau}\:{\rm e}^{\ii l\phi}P^{|k-l|}_j(\cos\Theta )\]
with $P^m_j(x)$, $-1\leq x\leq 1$, being an associated Legendre polynomial
defined by the following generating function:
\[(2m-1)!!{(1-x^2)^{{m\over 2}}t^m\over (1-2tx+t^2)^{m+{1\over
2}}}=\sum\limits_{j=m}^\infty t^jP^m_j(x).\]
Consequently, using the decomposed Laplacian the general solution can be 
written in the form 
\begin{equation}
f_0(r,\tau ,\phi ,\Theta
)=\sum\limits_{j=0}^\infty\sum\limits_{k,l=0}^j\lambda^{kl}_j\varrho^{kl}_j(r)
Y^{kl}_j(\tau, \phi ,\Theta )
\label{eloallitas}
\end{equation}
where $\lambda^{kl}_j$ are complex numbers and $\varrho^{kl}_j$ are
smooth radial functions. Notice that the spherical harmonics obey
\[\triangle_{S^3/\Z_N}Y^{kl}_j=-j(j+1)Y^{kl}_j,\:\:\:\:\:
{\partial\over\partial\tau} Y^{kl}_j=\ii kNY^{kl}_j\]
yielding that the solution of (\ref{laplace}) reduces to the ordinary
differential equations 
\begin{equation}
r^2{\dd^2\varrho^{kl}_j\over\dd r^2}+2r{\dd\varrho^{kl}_j\over\dd
r}-j(j+1)\varrho^{kl}_j=0
\label{radial1}
\end{equation}
if $c=0$ and
\begin{equation}
r^2{\dd^2\varrho^{kl}_j\over\dd r^2}+2r{\dd\varrho^{kl}_j\over\dd
r}-\left( j(j+1)+{r^2+4mNr\over 4m^2}k^2\right)\varrho^{kl}_j=0
\label{radial2}
\end{equation}
if $c=1$, $j=0,1,\dots$, $0\leq k,l\leq j$. From here we can see that
$\varrho^{kl}_j$ hence $f$ can have an isolated singularity only in the
origin of $\C^2/\Z_N.$

If (\ref{eloallitas}) is real and converges, by projecting the
corresponding Levi--Civit\'a connections of the rescaled metrics onto the
$\su^-$ part we can produce self-dual connections via the
Atiyah--Hitchin--Singer
theorem. To get {\it regular} solutions however, we have to propose
further regularity conditions on the rescaling harmonic functions. 
The Uhlenbeck theorem \cite{uhl} guarantees that generally in a given
gauge only {\it pointlike} singularities can be removed from
self-dual connections. Therefore we have to study the
singularities of the rescaled Levi--Civit\'a connections. The natural
orthonormal system to $g_V$, associated to the gauge (\ref{metrika}) is
\begin{equation}
\xi^0={1\over\sqrt{V}}(\dd\tau +\alpha
),\:\:\:\:\:\xi^1=\sqrt{V}\dd x
,\:\:\:\:\:\xi^2=\sqrt{V}\dd y,\:\:\:\:\:\xi^3=\sqrt{V}\dd z.
\label{bazis}
\end{equation}
Since $\tilde{g} =f^2g$ we have $\tilde{\xi}^i=f\xi^i$ consequently in
our gauge the Levi--Civit\'a connection $\tilde{\omega}$ with
respect to the rescaled metric obey the following Cartan equation on
$U_V$:
\[-\omega^i_j\wedge\xi^j+\dd (\log f)\wedge\xi^i =
-\tilde{\omega}^i_j\wedge\xi^j.\] 
As it is proved in \cite{ete-hau3} in this gauge the original connection
$\omega$ is self-dual hence cancels out if we project
$\widetilde{\omega}$ to the $\su^-$ component consequently all 
singularities of the self-dual connection $\nabla^-$ must come from
the $\dd (\log f)$ term in our gauge. Hence if we can understand the
singularities of the function $|\dd (\log f)|_{g_V}$ then
we can find the required regularity condition for the rescaling harmonic
functions. Notice that if $f$ is zero somewhere then $|\dd (\log
f)|_{g_V}$ diverges consequently {\it an acceptable harmonic function
must be positive and bounded everywhere, except isolated points.}

Taking into account that $\overline{Y^{kl}_j}=Y^{-k,
-l}_j$, a real basis for the solutions is given by
\[{1\over 2}(Y^{kl}_j+Y^{-k,-l}_j),\:\:\:\:\:{1\over
2\ii}(Y^{kl}_j-Y^{-k,-l}_j)\]
hence, via $\varrho_j^{kl}=\varrho^{-k,-l}_j$, if we have
$\overline{\lambda^{kl}_j}=\lambda_j^{-k, -l}$ we get real
solutions in (\ref{eloallitas}).

For $j=0$ (hence $k=l=0$) both (\ref{radial1}) and (\ref{radial2})
yields $c_1r^{-1}+c_2$.
Moreover $Y^{00}_0=1/4\pi$ hence separating the $j=0$ term in
(\ref{eloallitas}) we can write a harmonic function $f_0$ over
$(M_{V_0}, g_0)$ as
\[ f_0(r,\tau ,\phi ,\Theta )=c_1+{c_2\over r}+h_0(r,\tau ,\phi
,\Theta )\]
if $c=0,1$. The radial function $R^2=|v|^2+|w|^2=r^2+({\rm Im}\:w)^2$ in
$\C^2$ is invariant under $\Z_N$ hence we can talk about lense spaces of
radius $R$ centered at the origin of $\C^2/\Z_N$. Now take such a lense
space. By the aid of the explicit expressions for the
spherical harmonics it is not difficult to show that for $j>0$ 
\[\int\limits_{S^3/\Z_N}\left( Y^{kl}_j+Y^{-k,-l}_j\right)\dd
V_g={1\over 2\ii}\int\limits_{S^3/\Z_N}\left(
Y^{kl}_j-Y^{-k,-l}_j\right)\dd V_g=0\]
holds (integration is with respect to the induced metric on the lense
space in question) hence these real functions must change sign somewhere
on any lense space showing that in fact $h_0$ also changes sign i.e.
vanishes along three dimensional subsets in the above decomposition.
 
Furthermore the coefficients of (\ref{radial1}) and (\ref{radial2}) are
real analytic in $r$ hence any solution of them must be also real analytic
in $r$. In fact the general solution to (\ref{radial1}) and
(\ref{radial2}) with $k=l=0$ is $c_1r^{-j-1}+c_2r^j$ if $j\geq 0$. It is
also not difficult to see that any non-trivial solution of (\ref{radial2})
with $j>1$, $k,l\geq 0$ behaves either as $\varrho^{kl}_j(r)\sim r^{-j-1}$
if $r\rightarrow 0$ or $\varrho^{kl}_j(r)\sim r^j$ if
$r\rightarrow\infty$. This implies  that in both cases $h_0$ dominates
$f_0$ for small or large values of $r$ hence the total solution $f_0$ is
also plagued by small codimensional zero sets. For later
convenience we formulate this in the following
\begin{lemma}
Let $B^R_\varepsilon$ be the intersection of a ball of radius $R$ centered
at origin with the complement of a small ball of radius $\varepsilon$
also centered at the origin of $\C^2$. Let $U^R_\varepsilon
:=B^R_\varepsilon/\Z_N$ be its quotient. Then for a harmonic function
\[\lim\limits_\stack{R\rightarrow\infty}{\varepsilon\rightarrow
0}\:\inf\limits_{p\in U^R_\varepsilon}f_0(p)=-\infty,\:\:\:\:\:\lim\limits_
\stack{R\rightarrow\infty}{\varepsilon\rightarrow
0}\:\sup\limits_{p\in U^R_\varepsilon}f_0(p)=\infty\]
holds unless $h_0=0$ in its decomposition. $\Diamond$
\label{lemma}
\end{lemma} 
Consequently we must have 
\begin{equation}
f_0(r)=\lambda_0+{\lambda_1\over r}
\label{tnutmegoldas}
\end{equation}
with $0\leq\lambda_i\leq\infty$ for a generic acceptable harmonic
function on $(M_{V_0}, g_0)$ i.e., $f_0$ admits the full $U(2)$
symmetry. This solution was found in \cite{ete-hau2} and \cite{ete-hau3}
for the Taub--NUT metric. We remark that the  harmonic expansion on
Eguchi--Hanson space is treated in e.g. \cite{mal}.

Our aim is to find the appropriate generalization of the above classical
expansion (\ref{eloallitas}) in the general case by regarding the
Gibbons--Hawking metric as a perturbation of the singular metric $g_0$. To
achieve this, we have to construct the twistor spaces of the 
Gibbons--Hawking spaces.

\section{Regularization}

As it is well-known, Penrose' twistor theory, originally  
designed for flat Lorentzian $\R^4$ can be generalized to self-dual
four-manifolds. From our viewpoint this is important because the
Gibbons--Hawking spaces are self-dual (or half conformally flat) and by
the aid of their twistor spaces various differential equations can be
solved over them.

Let us recall the general theory \cite{bes}. Let $(M,g)$ be a
Riemannian spin four-manifold and consider the projective bundle
$Z:=P(S^-M)$ of the complex chiral spinor bundle $S^-M$ over it.
Clearly, $Z$ admits a fiber bundle structure $p : Z\rightarrow M$ with
$\C P^1$'s as fibers. By using the Levi--Civit\'a connection on $(M,g)$ we
can split the tangent space $T_zZ$ at any point $z\in Z$ and the
horizontal and vertical subspaces can be
endowed with natural complex structures hence $Z$, as a real six-manifold,
carries an almost complex structure, too. A basic theorem of Penrose or
Atiyah, Hitchin, Singer states that this complex structure is integrable,
i.e. $Z$ is a complex manifold if and only if $(M,g)$ is half conformally
flat (\cite{bes}, p.380, Th. 13.46). $Z$ is called the {\it twistor space}
of $(M,g)$. The fibers of $p : Z\rightarrow M$ are
then holomorphic projective lines in $Z$ with normal bundle $H\oplus H$
and belong to a family parameterized by a complex
four-manifold $\C M$. (We denote by $H^k$ ($k\in\Z$) the
powers of the tautological line bundle $H$ on $\C P^1$ while their induced
sheaves of holomorphic sections are $\co (k)$ respectively.) If one endows
$Z$ with a real structure $\tau : Z\rightarrow Z$
with $\tau^2 =$Id such that the fibers are kept fixed by the real
structure then $M\subset\C M$ is encoded in $Z$ as the family of
these ``real lines''. The basic example is $(M,g)=(S^4, \delta )$, the
round sphere, $\C M=Q^4$ the Klein quadric and $Z=\C P^3$.

If $g$ is Ricci flat then $S^-M$ with its induced connection is a
flat bundle. Hence if $M$ is moreover simply connected then $Z$ can be
contracted onto a particular fibre $\C P^1$ via parallel transport.
Consequently in this case we have another fibration $\pi
:Z\rightarrow\C P^1$ with fibers the copies of $M$. Hence if a holomorphic
line bundle $H^k$ is given on $\C P^1$ then it can be pulled
back to line bundles $\pi^*H^k$ over $Z$. Consider the sheaf
cohomology group $H^1(Z, \co (-2))$ of the line bundle $\pi^*H^{-2}$. Take
an element $[\omega ]\in H^1(Z, \co (-2))$ and consider for a representant
$\omega\in [\omega ]$ the complex valued real analytic function
$f:M\rightarrow \C$ given by the restriction $\omega\vert_{Y_p}$
where $\C P^1\cong Y_p\subset Z$ ($p\in M$) is a real line. Then we have
$\triangle f=0$ (with respect to $g$) and there is a natural bijection
\[T:H^1(Z, \co (-2))\cong{\rm ker}\:\triangle ,\]
onto real analytic soutions, called the {\it Penrose transform}. In
fact if the original metric $g$ is real analytic then this isomorphism
gives rise to {\it all} solutions to the Laplace equation in question. For
a proof and the more general statement see e.g. \cite{hit2}. Taking into
account the isomorphism $H^1(Y_p, \co (-2))\cong H^{(1,1)}(\C P^1)$ and
that $(1,1)$-forms can be integrated over $Y_p\cong\C P^1$ we can see that
the above transform is nothing but integration of forms over the real
lines, that is,
\[f(p)=\int\limits_{Y_p}\omega|_{Y_p}  .\]

In what follows we construct the above objects explicitly in case of
the Gibbons--Hawking spaces $(M_V, g_V)$ based on Hitchin's works.

{\it First consider the multi-Eguchi--Hanson spaces} i.e. the $c=0$ cases
\cite{hit1}. Since we have topologically $Z\cong
M_V\times\C P^1$ the twistor space $Z$ can be described analogouosly to
(\ref{ghter}) but instead of a single copy of $M_V$ we need a collection
of them, parameterized by a projective line. Therefore we have to replace
locally the constants $a_i$ with functions $a_i(u)$ depending
holomorphically on a complex variable $u\in\C P^1$. Hence, as a first 
approximation of $Z$ we define $\tilde{Z}$ by the equation
\[xy-(z^N+a_1(u)z^{N-1}+\dots +a_N(u))=0.\]
Since there are no nontrivial holomorphic functions on $\C P^1$ we have to
assume that $a_i$ is a holomorphic section of a bundle which implies that
the independent variables are also to be interpreted as elements of
various line bundles over $\C P^1$. The natural real structure
$\tau:Z\rightarrow Z$ 
\[\tau (x,y,z,u)=( (-1)^N\sigma (y), \sigma (x), -\sigma (z), \sigma
(u))\]
is induced by the $\Z_N$-invariant extension to $H^m\oplus H^n\oplus H^p$
of the natural antipodal map $\sigma$ on $\C P^1$ with $\sigma
(u)=-1/\overline{u}$. This map to be well-defined and the requirement that
the real lines have normal bundle $H\oplus H$ implies that $m=n=N$ and
$p=2$ and finally $\tilde{Z}$ is defined as a three dimensional complex
hypersurface $q(x,y,z,u)=0$ in the complex four-manifold $H^N\oplus
H^N\oplus H^2$ where $x,y\in H^N$, $z\in H^2$ and $a_i$ is an holomorphic
section of $H^{2i}$ obeying the algebraic equation 
\[q:H^N\oplus H^N\oplus H^2\longrightarrow H^{2N}\]
given by
\begin{equation}
q(x,y,z,u)=xy-\prod\limits_{s=1}^N(z-p_s(u)).
\label{Zegyenlet}
\end{equation}
Here we use the factorization
\[\sum\limits_{i=0}^Na_i(u)z^{N-i}=\prod\limits_{s=1}^N(z-p_s(u))\]
and $p_s$' are interpreted as holomorphic sections of $H^2$ in this
case. Unfortunately $\tilde{Z}$ is not a manifold because generically
there are points $u_{ij}^\pm\in\C P^1$ for which $p_i-p_j=0$ hence the
discriminant of $\sum\limits_{i=0}^Na_i(u)z^{N-i}$ vanishes. Consequently
the fiber over this point is singular. More precisely, there is a singular
point  $z^*:=(x=0, y=0, z=p_i(u_{ij}^\pm ), u_{ij}^\pm )$ on the fiber
over $u_{ij}^\pm$.

These singularities can be removed by deforming $\tilde{Z}$ within a
family of non-singular surfaces.
During the course of this deformation typically we have to take a branched
cover of $\C P^1$ over the singular points $u_{ij}^\pm$ hence the
resolved $Z$ would by first look fibered over a higher genus Riemann
surface. However in our special case we know {\it a priori} that the
twistor manifold $Z$ exists hence the resolved model is still fibered over
$\C P^1$. For our porposes it is enough to work with the singular model
$\tilde{Z}$.

Now we construct the real lines explicitly, following \cite{hit1}.
Remember that these are invariant sections of the holomorphic 
bundle $\pi :\tilde{Z}\rightarrow \C P^1$ with respect to the real
structure $\tau$. If we take $z_p(u)=au^2+2bu+c$ and
$p_s(u)=a_su^2+2b_su+c_s$ then the holomorphic sections $s_p:\C
P^1\rightarrow\tilde{Z}$ are given by polynomials $x_p(u), y_p(u)$ and
$z_p(u)$ obeying (\ref{Zegyenlet}). Consequently we ought to solve
it. However this equation can be solved by a simple
factorization yielding
\begin{equation}
x_p(u)=A\prod\limits_{s=1}^N(u-\alpha_{p_s}),\:\:\:\:\:
y_p(u)=B\prod\limits_{s=1}^N(u-\beta_{p_s}),\:\:\:\:\:z_p(u)=au^2+2bu+c
\label{EHszeles}
\end{equation}
where $AB=\prod\limits_{s=1}^N(a-a_s)$ and $\alpha_{p_s},\beta_{p_s}$ are
roots to the equations $z_p(u)-p_s(u)=0$.
The reality condition provided by the $\tau$-invariance gives $a =
-\overline{c}$, $b=\overline{b}$ and $a_s=-\overline{a}_s$,
$b_s=\overline{b}_s$ hence the roots explicitly look like
\begin{equation}
\alpha_{p_s}={-(b-b_s)-\sqrt{(b-b_s)^2+|a-a_s|^2}\over
a-a_s},\:\:\:\:\:\beta_{p_s}={-(b-b_s)+\sqrt{(b-b_s)^2+|a-a_s|^2}\over
a-a_s}.
\label{gyokok}
\end{equation}
The reality condition also yields
\begin{equation}
A={\rm e}^{\ii\phi}\sqrt{\prod\limits_{s=1}^N\left( b-b_s+
\sqrt{(b-b_s)^2+|a-a_s|^2}\right)}.
\label{ABegyenlet}
\end{equation}
(In these equations the square root is understood as the {\it positive} 
square root to avoid ambiguity.) From here we can see again that any real
section $s_p$ i.e. a point $p\in M_V$ away from the NUTs is parameterized
by $( {\rm Re}\:a, b, {\rm Im}\:a )\in\R^3$ and the angular variable
arg$A=:\phi\in S^1$.

Now we are ready to present integral formulae for the Laplacian over the
multi-Eguchi--Hanson spaces. Pick up a particular fibre $Y_p$ of $Z$, not
identical to any of the NUTs $Y_{p_s}$ and denote it by $\C P^1$. 
Take an $\omega$ with $[\omega ]\in H^1(Z\setminus\cup_s Y_{p_s},
\co (-2))$. Since this is holomorphic we can Taylor expand it. Denote
by $\xi$ and $\upsilon$ the tautological sections of the two copies of
$\pi^*H^N$ (as pulled back over $H^N$) and let $\zeta$ be the tautological
section of $\pi^*H^2$ (as pulled back over $H^2$). Hence we can write
\[\omega
=\sum\limits_{m,n,i}(\pi^*\omega_{mni})\xi^m\upsilon^n\zeta^i\]
where for $j:=Nm+Nn+2i$
\[[\omega_{mni}]\in H^1( \C P^1, \co (-2-j)),\:\:\:\:\:
\xi^m\upsilon^n\zeta^i\in H^0(H^j, \co (j)).\]
Since by definition the image of the tautological section is the zero
section of the corresponding line bundle, we can write over
$Z\setminus\cup_s Y_{p_s}$ that  
\[\xi = \xi -\pi^*x_p +\pi^*x_p,\:\:\:\:\:\upsilon =\upsilon -\pi^*y_p
+\pi^*y_p,\:\:\:\:\:\zeta =\zeta-\pi^*z_p +\pi^*z_p\]
and now $(\xi -\pi^*x_p)|_{Y_p}=0$ etc., i.e. they vanish exactly on the
image $Y_p$ of the section $s_p=(x_p, y_p, z_p)$. Consequently, by using
the isomorphism $s^*_p: H^0(Y_p , \co (j))\cong H^0(\C P^1, \co
(j))$ we can write
\[s_p^*(\omega |_{Y_p})=\sum\limits_{m,n,i} (s_p^*\pi^*\omega_{mni})
s_p^*((\xi -\pi^*x_p)|_{Y_p} +\pi^*x_p)^m((\upsilon -\pi^*y_p)|_{Y_p}
+\pi^*y_p)^n((\zeta-\pi^*z_p)|_{Y_p} +\pi^*z_p)^i\]
\[=\sum\limits_{m,n,i} \omega_{mni}x_p^my_p^nz_p^i.\]
Putting (\ref{EHszeles}) into the above expression
we find the following expansion for a harmonic function on the
multi-Eguchi--Hanson space with possible singularities in the NUTs:
\[f(p)=f({\rm Re}\:a,b,{\rm Im}\:a,{\rm arg}\:A)=\]
\begin{equation}
\sum\limits_{m,n,i}\:\int\limits_{\C
P^1}\omega_{mni}(u, \overline{u})A^{m-n}\prod_{s=1}^N
(a-a_s)^n\left(u-\alpha_{p_s}\right)^m\left(u-\beta_{p_s}\right)^n
(au^2+2bu- \overline{a})^i\dd u\wedge\dd\overline{u}.
\label{mEHmegoldas}
\end{equation}
where $\alpha_{p_s}$, $\beta_{p_s}$ are given by (\ref{gyokok}) and $A$
via (\ref{ABegyenlet}).

{\it Now we turn our attention to the case of multi-Taub--NUT spaces}. The
situation is similar \cite{che-kap1}\cite{hit3}. In this case the twistor
space topologically is the same as in the previous case but the complex
structure on it is different. This difference is realized by twisting the
previous line bundles with a new one. Consider the line bundle
$L^\mu (m)$ over $H^2$ defined as follows. Let $U_0(u\not=\infty ),
U_1(u\not= 0)$ be the standard covers of $\C P^1$ and
$\tilde{U}_0,\tilde{U}_1$ be the
induced covers of $H^2$. The coordinates on $\tilde{U}_0$ are $(u, z)$.
Suppose $H^m$ is the line bundle over $H^2$ pulled back from $\C P^1$. The
line bundle $L^\mu (m)$ over $H^2$ is defined by the transition function
($\mu$ is a parameter)
\[g(z,u):=u^{-m}{\rm e}^{-\mu z u^{-1}}\]
on $\tilde{U}_0\cap\tilde{U}_1$. Write $z=au^2+2bu-\overline{u}$
and consider a section $\eta^\mu$ of $L^\mu (N)$ as follows:
\[\eta^\mu (z,u):=\left\{\begin{array}{ll}
                 {\rm e}^{\mu ( au+b)} & \mbox{on
$\tilde{U}_0$},\\
g(z,u){\rm e}^{\mu ( au+b)} =u^{-N}{\rm
e}^{\mu (\overline{a}u^{-1}-b)} & \mbox{on $\tilde{U}_1$.}
\end{array}\right.\]
This $\eta^\mu$ is well defined because it is regular at both
$u=0$ and $u=\infty$ i.e., on $\tilde{U}_0$ and $\tilde{U}_1$. 
By using this section the space
$\tilde{Z}'$ is defined by \[\tilde{Z}':=\left\{ (x',y',z')\in
L^{1}(N)\oplus L^{-1}(N)\:\left|\right.
\:x'y'-\prod_{s=1}^N(z'-p_s(u))\right\}.\]
where the real sections are of the form
\[x'_p(u)=\eta^1(z_p(u),u)x_p(u),\:\:\:\:\:y'_p(u)=\eta^{-1}(z_p(u),u)y_p(u),
\:\:\:\:\:z'_p(u)=z_p(u)=au^2+2bu-\overline{a}.\]
Again, by the {\it a priori} knowledge of the existence of the
multi-Taub--NUT metric, this space can be deformed into a complex
analytic twistor space with holomorphic fibration $\pi ': Z'\rightarrow
\C P^1$. 

The corresponding integral formula, again with possible
singularities over the NUTs looks like
\[f(p)=f({\rm Re}\:a, b,{\rm Im}\:a,{\rm arg}\:A)=\]
\begin{equation}
\sum\limits_{m,n,i}\:\int\limits_{\C
P^1}\omega_{mni}(u, \overline{u})(A\eta )^{m-n}\prod_{s=1}^N
(a-a_s)^n\left(u-\alpha_{p_s}\right)^m\left(u-\beta_{p_s}\right)^n
(au^2+2bu-\overline{a})^i\dd u\wedge\dd\overline{u}
\label{mTNUTmegoldas}
\end{equation}
together with (\ref{gyokok}) and (\ref{ABegyenlet}). We may regard
(\ref{mEHmegoldas}) and (\ref{mTNUTmegoldas}) as generalizations of the
classical $U(2)$ symmetric harmonic expansion (\ref{eloallitas}) to the
general $U(1)$ symmetric case. Harmonic functions invariant under this
$U(1)$ isometry arise by taking $m=n$. In some sense a complementary
problem, i.e. the construction of the image of the Laplacian over
Gibbons--Hawking spaces was considered by constructing Green's 
functions in \cite{ati} and \cite{pag}.

After these preliminarities, we are ready to complete the analysis to find
the acceptable harmonic functions. Fix an $N$ and consider
the space $(M_V, g_V)$ with NUTs $p_1,\dots, p_N\in\R^3$. If $t\in [0,1]$
we can smoothly shrink the NUTs into the origin of $\R^3$ by taking
$a_s(t):=ta_s$, $b_s(t):=tb_s$ yielding a one-parameter family $(M_{V_t},
g_t)$ of Gibbons--Hawking spaces and associated Laplacians $\triangle_t$ 
which are regular for $t>0$ and blow down to the singular space
$(M_{V_0},g_0)$ with $\triangle_0$ if $t=0$.
Define the corresponding smooth functions
$\alpha_{p_s}(t),\beta_{p_s}(t)$ and $A(t)$ by (\ref{gyokok}) and
(\ref{ABegyenlet}) respectively and consider the family of harmonic
functions $f_t$ constructed this way via (\ref{mEHmegoldas}) and
(\ref{mTNUTmegoldas}). These satisfy $\triangle_tf_t=0$ and the flow
$t\mapsto f_t$ is {\it continuous} in the following sence. Note that for a
fixed $N$, the blow-up
map, restricted to the complement of the exceptional divisor $E$, induces
a biholomorphic mapping $\pi_t : M_{V_t}\setminus
E\rightarrow M_{V_0}$. Hence write $p$ for $\pi^{-1}_t(p)$ for all
$t>0$ and $p\in M_{V_0}$. Then there is a constant $C(p)>0$, such that
\[|f_t(p)-f_0(p)|\leq C(p)t.\] 
Moreover for fixed $c$ and $N$ this flow provides a 1:1 correspondence
between the solutions for $t>0$ and $t=0$; however the latter case is
explicitly known from the
previous section. There we have seen that the only acceptable solutions
are of the form (\ref{tnutmegoldas}) whose perturbation for $t>0 $ is (cf.
\cite{ete-hau3}) 
\begin{equation}
f_t(x)=\lambda_0+\sum\limits_{s=1}^N{\lambda_s\over |x-tp_s|}. 
\label{skala}
\end{equation}
All other harmonic functions $f_0$ are zero along three dimensional
subsets of $M_{V_0}$ close to either infinity or the
singularity. Consequently there is an
$\varepsilon >0$ such that if $0<t<\varepsilon$ their continuous
perturbation $f_t$ has the same property because of Lemma \ref{lemma}. 
 
Therefore we can conclude that a general acceptable harmonic
function on a Gibbons--Hawking space is of the form (\ref{skala}).
In the next section we summarize our findings. For this we introduce
notations, also used in \cite{ete-hau3}. Consider the quaternion-valued
1-form $\xi$ assigned to the basis (\ref{bazis}) and the imaginary
quaternion valued function assigned to (\ref{skala}) with $t=1$ (we denote
it by $f$) as follows:
\[\xi :=\xi^0+\xi^1\ii +\xi^2\jj +\xi^3\kk,\:\:\:\:\:\DD (\log f):=
{\partial\log f\over\partial x}\ii +{\partial\log f\over\partial y}\jj
+{\partial\log f\over\partial z}\kk.\]
\section{The theorems}
By collecting all the results from \cite{ete-hau2}, \cite{ete-hau3} and
the present work we can state:
\begin{theorem}
Let $(M_V, g_V)$ be a Gibbons--Hawking space with
$U_V:=M_V\setminus\{p_1,\dots ,p_N\}$ where $p_1,\dots,p_N$ denote the
NUTs of it. Then any smooth, finite action $SU(2)$ 't Hooft instanton
lives on the negative chiral bundle
$S^-M_V$ and such an $\nabla^-_{\lambda_0,\dots,\lambda_N}$ over this
space is given by \[A^-_{\lambda_0,\dots,\lambda_N}={\rm Im}\:{\DD (\log
f)\xi\over 2\sqrt{V}}\]
with $\nabla^-_{\lambda_0,\dots,\lambda_N}=\dd
+A^-_{\lambda_0,\dots,\lambda_N}$ in the gauge (\ref{bazis}) over $U_V$. 

The action of this connection is zero if $\lambda_s=0$ ($s>0$) otherwise
\[\Vert F^-_{\lambda_0,\dots,\lambda_N}\Vert^2=\left\{
          \begin{array}{ll}
         n-(1/N) & \mbox{if $\lambda_0=0$, $c=0$,} \\
                            &                             \\
           n & \mbox{otherwise,}
          \end{array}
                       \right. \]
where $N$ refers to the number of NUTs while $n$ is the
number of non-zero $\lambda_s$' ($s>0$).
$\Diamond$
\end{theorem}
All these instantons are $U(1)$ invariant hence dividing by this action we
have
\begin{theorem} 
Furthermore such an instanton corresponds to a $SU(2)$ magnetic
monopole $(\Phi ,A)$ over flat $\R^3$ where
\[\Phi ={\DD (\log f)\over 2},\:\:\:\:\:A ={\rm Im}\:{\DD (\log f)\ii\over
2}\dd x +{\rm Im}\:{\DD (\log f)\jj\over 2}\dd y+{\rm Im}\:{\DD (\log
f)\kk\over 2}\dd z\]
with singularities in the NUTs. All such magnetic monopoles have zero
magnetic charge.
$\Diamond$
\end{theorem}
In \cite{ete-hau3} we found all reducible connections and identified their
curvatures with $L^2$ harmonic 2-forms. Via a result in \cite{hau-hun-maz}
these generate the whole $L^2$ cohomology of $(M_V, g_V)$ hence:
\begin{theorem} 
A 't Hooft instanton $\nabla^-_{\lambda_0,\dots,\lambda_N}$ over $(M_V,
g_V)$ is reducible if and only if for an $s=0,\dots,N$ we have
$\lambda_s\not=0$
and $\lambda_r=0$ for $r=0,1,\dots,s-1,s+1,\dots,N$; in this case it can
be gauged into the form
\[B_s=\left( -{\dd\tau +\alpha \over
|x-p_s|V}+\alpha_s\right){\kk\over 2},\]
where $*_3\dd\alpha_s = \dd V_s$ with $V_s(x)=c+(2m/|x-p_s|)$.
The curvature $F_s$ of these connections generate the full $L^2$
cohomology of the Gibbons--Hawking space $(M_V, g_V)$.
$\Diamond$
\end{theorem}
\section{Concluding remarks}
In this paper we have classified all $SU(2)$ Yang--Mills instantons over
the Gibbons--Hawking spaces which arise by conformal rescalings of the the
metric (we called these instantons as 't Hooft instantons). During the
course of this we encountered the twistor manifolds of the
Gibbons--Hawking spaces which enables us to take an outlook.

Firstly at this point we remark that (\ref{mEHmegoldas}) and
(\ref{mTNUTmegoldas}) are generalizations of the classical formula of
Whittaker and Watson for solutions of the three dimensional Laplacian to
the Gibbons--Hawking spaces 
\cite{whi-wat}. This is because
dividing a Gibbons--Hawking space via its $U(1)$ isometry we recover the
flat $\R^3$. Indeed, taking $m=n$ in (\ref{mEHmegoldas}) or
(\ref{mTNUTmegoldas}) they cut down to
\[f(p)=\int\limits_{\C P^1}\sum\limits_{m,i}\omega_{mi}(u,\overline{u})
\prod_{s=1}^N(z_p(u)-p_s(u))^mz^i_p(u)\dd u\wedge\dd\overline{u}.\]
Pulling all NUTs into the origin of $\R^3$ we can write this as
\[f({\rm Re}\:a,b,{\rm Im}\:a)=\int\limits_{\C
P^1}\sum\limits_k\omega_k(u,\overline{u})(au^2+2bu-\overline{a})^k\dd
u\wedge\dd\overline{u}.\]
Writing $u=r{\rm e}^{\ii\phi}$ and performing the radial integration this
eventually yields 
\[f(x,y,z)=\int\limits_0^{2\pi} g(x\ii\cos\phi +y\ii\sin\phi
+z,\phi )\dd\phi\]
for some function $g$ which is the classical Whittaker--Watson formula
(also cf. \cite{mur}). 

Secondly, our method applies to more general gravitational instantons of 
Kronheimer \cite{kro2} as well as of Cherkis and Kapustin
\cite{che-kap1}\cite{che-kap2}. Taking into account that
$M_V\setminus\{p_1,\dots,p_N\}$ is connected and simply connected, by a
theorem of Buchdahl \cite{buc} we can see that 
\[H^1(Z\setminus\cup_sY_{p_s},\pi^*\co (-2-j))\cong H^1(\C P^1,\co
(-2-j))\]
and via Serre duality 
\[H^1(\C P^1,\co (-2-j))\cong H^0(\C P^1,\co (j))\cong\C^{1+j}.\]
Hence the $\pi^*(\omega_{mni})$ terms of our series indeed can be
interpreted as ``coefficients'' of a Taylor expansion yielding a
decomposition
\[H^1(Z\setminus\cup_sY_{p_s},\co 
(-2))\cong\bigoplus\limits_j\C^{1+j}.\]
Hence $H^1(Z\setminus\cup_sY_{p_s}, \co (-2))$ has an obvious element,
namely the pullback of one of the generators of $H^1(\C P^1 , \co
(-2))\cong\C$ i.e. the term corresponding to $j=0$. Exactly this element
gives the solutions (\ref{skala}) \cite{ati}\cite{hit2}. This
distinguished element exists because the twistor space $Z$ is
holomorphically fibered over $\C P^1$. However this is a consequence of
the fact that our gravitational instantons are not only self-dual but
Ricci-flat, too. Because the same is true for the new ``exotic'' ALE, ALF
and ALG spaces, without knowing the metric explicitly on them we can be
shure that 't Hooft instanton type solutions of the $SU(2)$ self-duality
equations exist over them. Identifying the reducible solutions as above we
can construct explicitly the $L^2$ cohomology of these spaces.

Finally we remark that we certainly cannot solve the self-dual $SU(2)$
Yang--Mills equations in their full generality with the conformal
rescaling method even in the simplest Gibbons--Hawking space. This
observations calls for
the generalization of the ADHM construction to all gravitational
instantons.
\vspace{0.1in}

{\bf Acknowledgement}. The author is grateful to S. Cherkis (IAS,
Priceton, USA) and T. Hausel (Univ. of Texas at Austin, USA) for the
useful conversations and calling our attention to further related
references.

\end{document}